\documentclass[journal]{IEEEtran}

\ifCLASSOPTIONcompsoc
  \usepackage[nocompress]{cite}
\else
  \usepackage{cite}
\fi

\usepackage[utf8]{inputenc}
\usepackage{chngcntr}
\usepackage[cmex10]{amsmath}
\usepackage{hyperref}
\usepackage[T1]{fontenc}
\usepackage{lettrine}
\usepackage{amsfonts}
\usepackage{amssymb}
\usepackage{graphicx}
\usepackage{csquotes}
\usepackage{mathtools}
\usepackage{bbm}
\usepackage[normalem]{ulem}
\usepackage{tabularx}
\usepackage[nottoc]{tocbibind}
\usepackage{float}
\usepackage[caption = false]{subfig}
\usepackage{enumerate}
\usepackage[dvipsnames]{xcolor}
\usepackage{tikz}
\usepackage{pgfplots}
\usepackage{physics}
\usepackage{todonotes}
\tikzset{>=latex}
\usepackage{blochsphere}
\usepackage{algorithm}
\usepackage[noend]{algpseudocode}
\algnewcommand\algorithmicforeach{\textbf{for each}}
\algdef{S}[FOR]{ForEach}[1]{\algorithmicforeach\ #1\ \algorithmicdo}

\usepackage{textgreek}
\usepackage[binary-units=true]{siunitx}

\makeatletter
\def\ps@headings{%
\def\@oddhead{\small{ACCEPTED FOR PUBLICATION IN IEEE COMMUNICATION LETTERS}\hfil \small{\thepage}}%
\def\@evenhead{\small{ACCEPTED FOR PUBLICATION IN IEEE COMMUNICATIONS LETTERS} \hfil \small{\thepage}}%
}
\makeatother

\usepackage{bytefield}
\usetikzlibrary{shapes, arrows}

\usepackage[shortcuts,acronym]{glossaries}
\newacronym{gewi}{GEWI}{Generate Entanglement While Idle}
\newacronym{qegp}{QEGP}{Quantum Entanglement Generation Protocol}
\newacronym{fifo}{FIFO}{First-In-First-Out}

\makeatletter

\newcounter{protocol}
\newenvironment{protocol}[1][htb]{%
  \let\c@algorithm\c@protocol
  \renewcommand{\ALG@name}{Protocol}
  \begin{algorithm}[#1]%
  }{\end{algorithm}
}

\title{Integrating Quantum Simulation for Quantum-Enhanced Classical Network Emulation}

\author{Stephen~DiAdamo,
        Janis~N\"otzel,
        Simon~Sekav\v{c}nik, 
        Riccardo~Bassoli,~\IEEEmembership{Member,~IEEE,}
        Roberto~Ferrara, 
        Christian~Deppe,~\IEEEmembership{Member,~IEEE,} 
        Frank~H.P.~Fitzek,~\IEEEmembership{Senior Member,~IEEE,}
        Holger~Boche,~\IEEEmembership{Fellow,~IEEE}\vspace{-6mm}%
\thanks{S. DiAdamo, J. N\"otzel, S. Sekav\v{c}nik, R. Ferrara, C. Deppe and H. Boche are with the Department of Electrical and Computer Engineering at Technische Universität München, München, Germany. H. Boche and J. N\"otzel are also with the Munich Center for Quantum Science and Technology (MCQST), München, Germany.
(e-mail: \{stephen.diadamo, janis.noetzel, simon.sekavcnik, roberto.ferrara, christian.deppe, boche\}@tum.de).}
\thanks{R. Bassoli and F. H.P. Fitzek are with the Deutsche Telekom Chair of Communication Networks, Institute of Communication Technology, Faculty of Electrical and Computer Engineering, Technische Universität Dresden, Dresden, Germany. F. H.P. Fitzek is also with the Centre for Tactile Internet with Human-in-the-Loop (CeTi), Cluster of Excellence, Dresden, Germany.
(e-mail: \{riccardo.bassoli, frank.fitzek\}@tu-dresden.de).}
\thanks{ \raisebox{0.4mm}{\textcopyright}~2021 IEEE. Personal use of this material is permitted. Permission from IEEE must be obtained for all other uses, in any current
or future media, including reprinting/republishing this material for
advertising or promotional purposes, creating new collective works,
for resale or redistribution to servers or lists, or reuse of any
copyrighted component of this work in other works.}
}

\pgfplotsset{compat=1.15}
\begin{document}
\maketitle
\bstctlcite{IEEEexample:BSTcontrol}

\begin{abstract}
We describe a method of investigating the near-term potential of quantum communication technology for communication networks from the perspective of current networks. For this, we integrate an instance of the quantum network simulator QuNetSim at the link layer into the communication network emulator ComNetsEmu. This novel augmented version of ComNetsEmu is thereby enabled to run arbitrary quantum protocols between any directly connected pair of network hosts. To give an example of the proposed method, we implement the link layer method of generating and storing entanglement while idle, to accelerate data transmission at later times using superdense coding.
\end{abstract}


\thispagestyle{headings}

\section{Introduction}
 
\IEEEPARstart{F}{uture} communication networks will encounter performance boundaries that will be difficult to overcome using traditional approaches. In fact, significant trade-offs will make requirements for very-high data rates (at very-low latency with extreme reliability and resilience) very difficult. At the same time, networks need to continue implementing complete network softwarization and massive data mining and processing because of in-network Artificial Intelligence (AI) \cite{CompBook}, for example. To go beyond these technological limitations, new communication paradigms are needed. Recent advances in entanglement-assisted data transmission \cite{guha2020infinite} highlight the potential of quantum communication methods for hybrid networks, utilizing these novel methods at lower network layers: this advantage being simplified by the integration with the existing infrastructure \cite{notzel2020entanglement}. The question arises whether this concept of hybrid classical-quantum networks carries further than described previously \cite{notzel2020entanglement,notzel2020entcomm,notzel2020-entanglement-enabled}. As the answer to this question is of a highly interdisciplinary nature, with this article we highlight a full integration of the quantum network simulator QuNetSim \cite{DiadamoNozelZangerMert2020} into the link layer of the classical network emulator ComNetsEmu \cite{CompBook}. With this tool, a variety of networking questions can be answered using the large set of possible quantum communication techniques \cite{QuantumBook2021} as part of hybrid classical-quantum communication networks. 

We display the potential use of the software integration with a clean focus on the case of a hybrid entanglement-assisted quantum-classical communication network. We assume that the protocol stack above the link layer remains unchanged. The link and the physical layer, however, are modified to enable quantum information processing and communication. In addition to the integration, we propose a novel link-layer protocol, which implements the queuing model in \cite{notzel2020entanglement}, where entanglement-assisted data transmission is a stochastic process generating, sharing, and storing entanglement between network nodes during stagnant communication periods, enabling higher data rates. If the stored entanglement is depleted during data transmissions, any further message in the buffer of the sending node is transmitted using classical communications. This protocol is labeled \enquote{\acrlong{gewi}} (\acrshort{gewi}, see \cite{notzel2020entanglement}), and we use it as a proof-of-concept for our emulation tool. Finally the protocol is evaluated over a single network link using the integrated software platform. In this way, we close an existing gap between quantum network simulation and classical network emulation. 

\section{Related Work}
In this section we review the related work from the areas of quantum network simulation and quantum link layer protocols.

\subsection{Quantum Communication Network Simulation}
The area of hybrid classical-quantum networks is a novel domain, with many open challenges remaining. To the best of our knowledge, no tool exists yet with the ability to emulate a classical network with additional quantum technological features. However, in the domain of quantum network simulation platforms, there have been many very recent developments  \cite{DiadamoNozelZangerMert2020,dahlberg2018simulaqron,coopmans2021netsquid,wu2020sequence,quisp}. A detailed overview of these platforms can be found in \cite{DiadamoNozelZangerMert2020} and \cite[Chapter~6.5]{QuantumBook2021}. The available quantum network simulators are developed for a variety of use-cases and use various simulation methodologies as well. Here, we consider those which use real-time and not discrete-event simulation. The reason for this is that much of the challenge of programming the logic that reacts to events originating in another system---in this case ComNetsEmu---is handled with a real-time simulator. Real-time network simulations do not necessarily terminate when there are no events pending in the simulation, and the CPU resources for keeping the program alive are minimal. Regarding the quantum network simulators that use real-time simulation, there are QuNetSim and SimulaQron \cite{dahlberg2018simulaqron}. SimulaQron is a quantum network simulator used for developing applications and runs in a multi-processed environment. QuNetSim runs in a multi-threaded environment in a single process and has many of the same features of SimulaQron, with the additional features that come from its design principle of mimicking the idle periods in a network. In contrast to SimulaQron, this brings the feature that nodes in the network are by default always running idly, waiting for incoming events to process. Thus, in QuNetSim, by default, one has the ability to program the network nodes to react---in real-time--- to incoming events generated in ComNetsEmu, making this software a suitable choice for the integration. An additional feature of QuNetSim is the control over how qubits that are transmitted are accessed at the receiving node. In QuNetSim, the arriving qubits are automatically stored at the receiving node, and the receiving node can then act on the stored qubits while more qubits continue to arrive. As far as we can tell, such a feature would need to be developed as an additional layer in SimulaQron. Overall, this makes QuNetSim a more suitable choice for the type of integration into a network emulator, as it is described here.

\subsection{Link Layer Protocols for Quantum Networks}

The following briefly reviews existing link-layer protocols for quantum networks and explains some important concepts for understanding the contribution of this article. In this work, the link layer is considered as a mechanism providing error-free qubit transmission between the interfaces connecting the network nodes. A network not making use of quantum effects will encode one bit per qubit at those interfaces. A network utilizing quantum effects can, in addition to this, distribute, store, and measure Einstein-Podolsky-Rosen (EPR) pairs between any two connected interfaces and execute arbitrary gates on each interface. A complete overview of this process is found in  \cite[Chapter~6.4]{QuantumBook2021}. This encapsulation of quantum functionalities into the lower layers simplifies the integration in terms of software and hardware, and is also in line with recent research, which we review in the following.

In \cite{notzel2020entcomm}, a classical-quantum network layer is considered, which can increase network throughput using entanglement-assisted message transmission. Features for a link-layer protocol are also proposed, which aim to generate entanglement for subsequent use in entanglement-assisted message transmission. Missing from~\cite{notzel2020entcomm} is the explicit protocol used on the link layer to transmit classical messages using the quantum channel. It only considers the statistical properties of the link layer to measure the performance of the network. 
    
Pirker and D\"ur propose a protocol stack for a quantum network~\cite{pirker2019quantum}. They introduce a so-called \enquote{connectivity layer} between the quantum physical and link layer. Such a layer does not have an analog in the Open Systems Interconnection (OSI) model \cite{zimmermann1980osi}. The connectivity layer handles the requests from upper layers by converting them into instructions for the physical layer. Such instructions can, for example, be a request for qubit transmission or EPR pair generation, without assumptions of the underlying physical models. Moreover, the connectivity layer provides means of handling quantum errors arising at the physical layer, detaching the link-layer logic from the physical layer. The proposed link layer further keeps track of the current quantum entanglement status in the network in order to later generate long-distance quantum entanglement. The connectivity layer hides the particular implementation of the physical layer from the higher layers. Our exemplary implementation of a network link, utilizing superdense coding, rests on the assumption of error-free transmission of (entangled) qubits, a functionality that can be achieved using functionalities of a connectivity layer.

In~\cite{dahlberg2019link}, a quantum network link-layer protocol is proposed for the generation and distribution of entanglement between network nodes. The main focus in the work is establishing multi-party entanglement in quantum networks rather than a mixture of entanglement generation and classical communication. The duty of the proposed link layer is to schedule which requests for entanglement will be served first. In the work, this is left open to the specific application and so one can consider the \acrshort{gewi} protocol as an explicit schedule to the proposed protocol. Moreover, since here we use the quantum channel to transmit classical information as well as entanglement, one could make use of the \enquote{Create and Keep} protocol to store entanglement to perform entanglement-assisted communication.

Among these works, only \cite{notzel2020entcomm} explicitly considers enhancing communication rates in classical networks with entanglement, but an explicit protocol is not considered. The use of entanglement in the other works is primarily for entanglement swapping routines or teleportation of quantum states. Using entanglement primarily for enhancing classical communication rates, and generating entanglement only when the network is idle, allows the network to see performance improvements when entanglement is established, and still function normally otherwise. We consider both the explicit link-layer protocol as well as a simulation tool for determining these performance improvements.

\section{Assumptions Made in the Integration}

Using an entanglement-assisted classical-quantum network requires new protocols to make use of the additional abilities of the network. We assume: 1) At each interface between a communication link and a network node, qubits can be stored in and retrieved from a quantum memory; 2) In each interface, EPR pairs can be generated with a heralding success signal and transmitted to the corresponding interface at another node via an error-free quantum channel; 3) Each interface in the network has the ability to perform a controlled Pauli-X gate (CNOT), Bell measurements, and the Pauli gates on qubits; 4) Interfaces can generate and transmit qubits on demand; 5) Joint measurements at each interface can distinguish the four EPR pairs; 6) Each interface can forward classical messages to other interfaces in the same node. A possible way of achieving these functionalities in a practical demonstration is by using the connectivity and link layers proposed in \cite{pirker2019quantum}. The classical upper layers remain unchanged. 

With these assumptions, the link layer can transmit a data frame of qubits such that the receiver can interpret whether to measure data or store EPR pairs. The link layer protocol on the receiver side can also reconstruct the classical information packet. The assumptions are tailored such that a) problems at the physical layer are abstracted away from the integration, b) the creation of complex quantum states via the network is limited such that simulations remain feasible c) the maximum flexibility is provided between network interfaces. 

Thus the link layer resulting from our integration is capable of demonstrating the \acrshort{gewi} protocol. For this protocol to work, long qubit storage times are required. The recent progress in this domain shows store-and-retrieve rates of 90\% for \SI{10}{\us} for 225 individually accessible memory cells, with a proposal to extend the duration into the seconds range \cite{pu2017experimental}, which are in a range that can be useful to benefit from variations in classical traffic load. 

This first implementation of a proof-of-concept of a quantum network simulator integrated into a classical emulator trades simplicity and flexibility for practicality: the true benefits of entanglement-assisted data transmission can be expected to arise in very specific transmission systems exploiting quantum effects right at the physical layer \cite{guha2020infinite}, and these applications do not necessitate a conversion into qubits as achieved via \cite{pirker2019quantum} which is used in our initial integration.

\section{Simulation Setup and Configuration}
    
The simulation is an integration of a quantum link layer and physical layer simulator offered by QuNetSim \cite{DiadamoNozelZangerMert2020}, with the classical network emulator ComNetsEmu \cite{CompBook}. The code for this work can be found at \cite{Sekavcnik_QontainerNet_2021}. From an implementation viewpoint, the quantum lower layers act as a \enquote{man-in-the-middle} between the communicating network nodes. This software entity intercepts network packets en route and converts the incoming data to binary data, as it would for any link layer. Using the binary format, a qubit data frame is constructed and transmitted using QuNetSim, with which we model a quantum point-to-point channel. The end-nodes of this channel perform the quantum tasks, such as the qubit encoding, transmission, storage, retrieval, and decoding. The two end-nodes run in a multi-threaded QuNetSim application, which can run any communication protocol. 

To integrate ComNetsEmu and the QuNetSim simulation, a bridge interface is used, responsible for moving traffic generated in ComNetsEmu to the QuNetSim application and back again. With ComNetsEmu, the quantum simulation part of the program is \enquote{containerized} in a Docker environment. Docker uses \enquote{containers} to isolate code and all its dependencies in virtual computing environments \cite{CompBook}. ComNetsEmu simplifies the deployment of containers specifically, to simulate communication networks. One can configure the IP addresses of these containers and monitor the incoming and outgoing traffic flow of the container. Using ComNetsEmu, the bridge interface is deployed in a container, which runs QuNetSim as an application and monitors the network traffic like a \enquote{man-in-the-middle}. 

The bridge is configured such that it routes all the network-layer traffic to a particular queue and it is configured to monitor the incoming traffic arriving in this queue. When packets arrive, they are disassembled into binary strings and processed in the QuNetSim application. The quantum channel running in the QuNetSim application converts the binary strings into data frames and transmits them over the channel to the QuNetSim Host, representing the receiver's link layer. Within the QuNetSim application, the destination receives the qubits and decodes the frames according to the particular algorithm and reconstructs the binary string once the end of the data frame is detected. Once completed, the network-layer packet is reassembled and sent to the bridge to proceed with routing.

\section{Proof-of-Concept: A Link-Layer Protocol for Classical-Quantum Communication Networks}
    
In this section, we describe a link-layer protocol to perform entanglement-assisted transmissions which we will then use with the classical-quantum simulation platform described in the previous section for proof-of-concept. The protocols referred to are defined below. Protocol \ref{proto:link-layer} contains the heart of the logic. In each iteration of Protocol \ref{proto:link-layer}, the sender checks if there is classical data to send at the upper layers. If there is data, then a quantum data frame is generated and transmitted using Algorithm~\ref{algo:send_data}, which uses the state of the entanglement buffer to determine if the classical data frame should be encoded into qubits (thus encoded into an entanglement-assisted data frame) or not, using the stored EPR pairs in a \acrfull{fifo} ordering. On the other hand, if there is no data to transmit, a frame of EPR qubits is sent if the quantum storage can accommodate it.

\begin{figure}[ht]
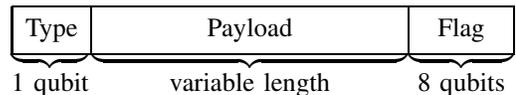

	\centering
	\begin{bytefield}[bitwidth=1.0em]{28}
		\bitbox{3}{Type} & 
		\bitbox{12}{Payload} & 
		\bitbox{4}{Flag} \\
		\bitbox[]{3}{$\hspace{-0.4mm}\underbrace{\hspace{1.03cm}}_{\text{\normalsize 1 qubit}}$}
		\bitbox[]{12}{$\underbrace{\hspace{4.2cm}}_{\text{\normalsize variable length}}$}
		\bitbox[]{4}{$\underbrace{\hspace{1.4cm}}_{\text{\normalsize 8 qubits}}$}
	\end{bytefield}
	\caption{The link-layer data frame used for the entanglement-assisted quantum channel.}
	\label{fig:data_frame}
\end{figure}
    
The data frame structure we use is depicted in Fig.~\ref{fig:data_frame}. One leading qubit is used to indicate the \emph{Type} of payload---entanglement or data---followed by the payload itself, which is of variable length. The benefit of this is the avoidance of padding, which could result in a waste of quantum resources. Next, a reserved byte sequence \emph{Flag}, representing the \enquote{end of frame}, signals the end of the frame. Alternatively, we could have included the number of bits transmitted in the frame, but when this binary number is greater than 255 bits, it becomes less efficient than using a termination flag.  When an EPR frame is sent, the protocol uses a maximum size of the EPR payload $L$. In this noiseless and ideal setting, we exclude a leading flag for the data frame, since qubits here are a discrete resource, for example, a single polarized photon. After the end of a sequence, any qubit detection would indicate the beginning of a new incoming frame. In more realistic scenarios, the data frame would need to be modified according to the hardware parameters.

Once a data frame is received, the receiver performs Algorithm~\ref{algo:receive_frame}. The receiver simply decodes the first incoming qubit to determine the payload type. If the header indicates EPR pairs, they store the payload for later use. If the header indicates data, then the information is decoded either with entanglement-assistance or without. The sender and receiver agree with this before Protocol \ref{proto:link-layer} starts.     

\begin{protocol}
	\caption{Entanglement-Assisted Message Transmission}
	\small{
	\textbf{Sender}
	\begin{algorithmic}[1]
		\While {\it{active}}
		\If {\it{there is data to transmit}}
		\it{send data frame}
		\ElsIf {\it{entanglement buffer is not full}}
		\it{send EPR frame}
		\EndIf
		\EndWhile
	\end{algorithmic}
	        
	\textbf{Receiver}
	\begin{algorithmic}[1]
		\While {\it{active}}
		\it{receive frames}
		\EndWhile
	\end{algorithmic}}
	\label{proto:link-layer}
\end{protocol}
\begin{algorithm}[ht]
	\caption{Send Data Frame}
	\small{
	\begin{algorithmic}[1]
		\State $payloadQubits \gets$ \it{initialize empty list}
		\State $q_h \gets$ \it{initialize  qubit in the $\ket{1}$ state}
		\For{byte \textbf{in} data stream}
		    \For {$(b_1$, $b_2)$ \textbf{in} byte} \Comment{iterate 2 bits at a time}
		        \If {\it{ent. buffer non-empty}}
		            \State $q_1 \gets$ \it{pop qubit from ent. buffer}
		            \State \it{superdense encode} $q_1$ with $(b_1, b_2)$
		            \State \it{add $q_1$ to} $payloadQubits$
	            \Else
		            \State $q_1, q_2\gets$ \it{initialize two qubits in $\ket{0}$ state}
		            \If {$b_1=1$} $X(q_1)$ \Comment{excite $q_1$ to $\ket{1}$} \EndIf 
		            \If {$b_2=1$} $X(q_2)$ \Comment{excite $q_2$ to $\ket{1}$} \EndIf 
		            \State \it{add $q_1, q_2$ to} $payloadQubits$
		        \EndIf
	        \EndFor
	    \EndFor
	    \State $flagQubits \gets$ \it{generate 8 qubit flag state}
		\State $frame \gets (q_h, payloadQubits, flagQubits)$
		\State \it{send frame}
	\end{algorithmic}\label{algo:send_data}}
\end{algorithm}
\begin{algorithm}[ht]
	\caption{Send EPR Frame}
	\small{
	\begin{algorithmic}[1]
		\State $payloadQubits \gets$ \it{initialize empty list}
		\State $q_h \gets$ \it{initialize qubit in the $\ket{0}$ state}
		\While{ent. buffer not full \textbf{and}  $|payloadQubits| < L$}
		\State $q_{1}, q_{2} \gets$ \it{initialize two qubits in $\ket{0}$ state} 
		\State \it{entangle} $q_{1}, q_{2}$
		\State \it{store} $q_{1}$ in ent. buffer
		\State \it{add} $q_{2}$ to $payloadQubits$
		\EndWhile
		\State $flagQubits \gets$ \it{generate 8 qubit flag state}
		\State $frame \gets (q_h, payloadQubits, flagQubits)$
		\State \it{send frame}
	\end{algorithmic}\label{algo:send_epr}}
\end{algorithm}
\begin{algorithm}[ht]
	\caption{Receive Frame}
	\small{
	\begin{algorithmic}[1]
		\State $m_h \gets$ \it{receive header qubit and measure}
		\State $bytes \gets$ \it{initialize empty list} \State $bits \gets$ \it{initialize empty list}
		\If {$m_h = 1$} \Comment{Data frame}
		\While{$bits \neq FLAG$}
		\State \it{reset $bits$}
		    \While {$|bits| \neq 8$}
		        \State $q \gets $ \it{receive qubit}
        		\If{\it{ent. buffer non-empty}}
        		\State $q_e \gets$ \it{pop qubit from ent. buffer}
        		\State $b_0, b_1 \gets$ \it{superdense decode $(q, q_{e}$})
        		\State \it{add} $(b_0, b_1)$ to $bits$
        		\Else \Comment{classical decode}
        		\State $b_0 \gets measure(q)$ 
        		\State $bits.add(b_0)$
        		\EndIf
		    \EndWhile
		\State add $bits$ to $bytes$
		\EndWhile
	    
		\Else \Comment{EPR frame}
		\State $qubits \gets$ \it{receive $L$ qubits}
		\State \it{add} $qubits$ to ent. buff
		\EndIf
	\end{algorithmic}\label{algo:receive_frame}}
\end{algorithm}

\section{Simulations of Bursty Network Traffic with Entanglement Assistance}

\begin{figure}[ht]
    \centering
    \begin{tikzpicture}[
  declare function={
    func(\b,\d,\e,\l) = (\e * \l <= (\b*\d / 2)) * ((\b*\d)/(\b*\d - \e*\l) )   +
              (\e * \l > \b*\d/2) * (2)
   ;
  }
]

\begin{axis}
[
    xlabel={EPR frame size (8 qubits)}, 
    ylabel={Avg. bits per data qubit},
    ylabel style={at={(axis description cs:-0.08,.5)}},
    label style={font=\small},
    grid, 
    xmin=0, 
    ymin=0.50, 
    ymax=2.15,
    height=4.75cm,
    width=8.75cm,
    cycle list name=exotic,
    domain=0:14,
    samples=200,
    legend pos=south east,
    legend style= {font=\footnotesize},
    legend columns = 2,
    legend cell align=left,
]

\addlegendimage{mark=*, mark size=1.5, dashed, thick, Red}
\addlegendimage{mark=*, mark size=1.5, dashed, thick, BurntOrange}
\addlegendimage{mark=*, mark size=1.5, dashed, thick, Green}
\addlegendimage{mark=*, mark size=1.5, dashed, thick, MidnightBlue}

\addplot[mark=*, only marks, mark options={fill=Red}, mark size=1.5, Red]
table[col sep=space] {%
0 1
1 1.909090909
2 2
3 2
4 2
5 2
6 2
7 2
8 2
9 2
10 2
11 2
12 2
13 2
};
\addplot[dashed, thick, Red, forget plot]
{
func(1, 168, 10, x*8)
};
\addlegendentry{1 Packet Bursts}

\addplot[mark=*, only marks, mark options={fill=BurntOrange}, mark size=1.5, BurntOrange]
table {%
0 1
1 1.209424084
2 1.483146067
3 1.913043478
4 2
5 2
6 2
7 2
8 2
9 2
10 2
11 2
12 2
13 2
};
\addplot[dashed, thick, BurntOrange, forget plot]
{
func(3, 168, 10, x*8)
};
\addlegendentry{3 Packet Bursts}

\addplot[mark=*, only marks, mark options={fill=Green}, mark size=1.5, Green]
table {%
0 1
1 1.11627907
2 1.263157895
3 1.426751592
4 1.635036496
5 1.914529915
6 2
7 2
8 2
9 2
10 2
11 2
12 2
13 2
};
\addplot[dashed, thick, Green, forget plot]
{
func(5, 168, 10, x*8)
};
\addlegendentry{5 Packet Bursts}

\addplot[mark=*, only marks, mark options={fill=MidnightBlue}, mark size=1.5, MidnightBlue]
table {%
0 1
1 1.05
2 1.105263158
3 1.166666667
4 1.235294118
5 1.3125
6 1.4
7 1.5
8 1.615384615
9 1.75
10 1.909090909
11 2
12 2
13 2
};
\addplot[dashed, thick, MidnightBlue,  forget plot]
{
func(10, 168, 10, x*8)
};
\addlegendentry[solid]{10 Packet Bursts}

\end{axis};
\draw[fill=white, draw=black] (5, 2.0) rectangle (6.95, 1.15);
\node (a) at (5.3, 1.7) {$\Large{\boldsymbol{\cdot}}$};
\node (b) at (6.2, 1.76) {\footnotesize{Simulation}};
\node () at (6.125, 1.35) {\footnotesize{Analytical}};
\draw[dashed, thick] (5.1, 1.35) --  (5.4, 1.35);

\end{tikzpicture}
    \begin{tikzpicture}[
  declare function={
    func(\b,\d,\e,\l) = (\e * \l <= (\b*\d / 2)) * ((\b*\d)/(\b*\d - \e*\l) )   +
              (\e * \l > \b*\d/2) * (2)
   ;
  }
]

\begin{axis}
[
    xlabel={Number of EPR packets sent between bursts}, 
    ylabel={Avg. bits per data qubit}, 
    ylabel style={at={(axis description cs:-0.08,.5)}},
    label style={font=\small},
    grid, 
    xmin=0, 
    ymin=0.50, 
    ymax=2.15,
    height=4.75cm,
    width=8.75cm,
    legend pos=south east,
    cycle list name=exotic,
    domain=0:8,
    samples=200,
    legend style= {font=\footnotesize},
    legend columns = 2,
    legend cell align=left,
]

\addlegendimage{mark=*, mark size=1.5, dashed, thick, Red}
\addlegendimage{mark=*, mark size=1.5, dashed, thick, BurntOrange}
\addlegendimage{mark=*, mark size=1.5, dashed, thick, Green}
\addlegendimage{mark=*, mark size=1.5, dashed, thick, MidnightBlue}

\addplot[mark=*, only marks, mark options={fill=Red}, mark size=1.5, Red]
table {%
0 1
1 2
2 2
3 2
4 2
5 2
6 2
7 2
8 2
};

\addlegendentry{1 Packet Bursts}
\addplot[dashed, thick, Red, forget plot]
{
func(1, 168, x, 20*8)
};

\addplot[mark=*, only marks, mark options={fill=BurntOrange}, mark size=1.5, BurntOrange]
table {%
0 1
1 1.4651
2 2
3 2
4 2
5 2
6 2
7 2
8 2
};
\addplot[dashed, thick, BurntOrange, forget plot]
{
func(3, 168, x, 20*8)
};
\addlegendentry{3 Packet Bursts}

\addplot[mark=*, only marks, mark options={fill=Green}, mark size=1.5, Green, forget plot]
table {%
0 1
1 1.2388
2 1.62
3 2
4 2
5 2
6 2
7 2
8 2
};
\addlegendentry{5 Packet Bursts}
\addplot[dashed, thick, Green]
{
func(5, 168, x, 20*8)
};

\addplot[mark=*, only marks, mark options={fill=MidnightBlue}, mark size=1.5, MidnightBlue,  forget plot]
table {%
0 1
1 1.1052
2 1.2564
3 1.42216
4 1.6153
5 1.9182
6 2
7 2
8 2
};
\addlegendentry{10 Packet Bursts}
\addplot[dashed, thick, MidnightBlue]
{
func(10, 168, x, 20*8)
};

\end{axis};

\draw[fill=white, draw=black] (5, 2.0) rectangle (6.95, 1.15);
\node (a) at (5.3, 1.7) {$\Large{\boldsymbol{\cdot}}$};
\node (b) at (6.2, 1.76) {\footnotesize{Simulation}};
\node () at (6.125, 1.35) {\footnotesize{Analytical}};
\draw[dashed, thick] (5.1, 1.35) --  (5.4, 1.35);

\end{tikzpicture}
    \caption{Upper: Bursty traffic with varying EPR frame length with 10 EPR frames between bursts using superdense coding. Lower: Bursty traffic with varying number of EPR frames between 168-bit packet bursts with 160-qubit EPR frames using superdense coding.}
    \label{fig:data_1}
\end{figure}
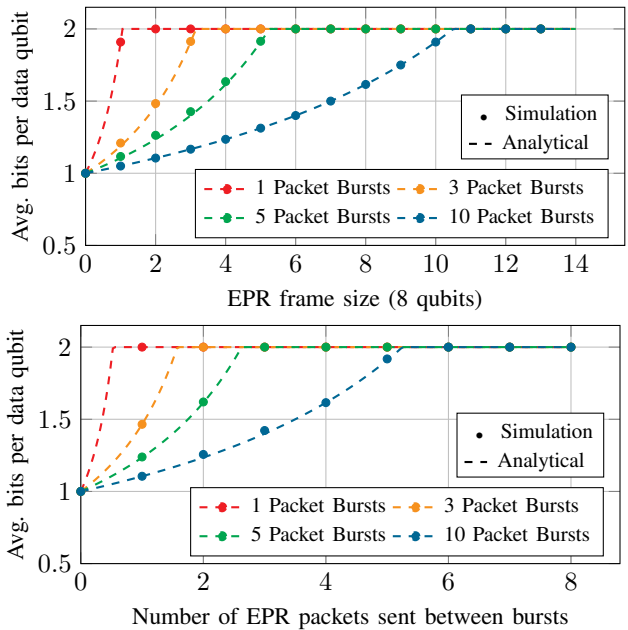

As a proof of concept using the simulation framework for quantum-enhanced classical networks, we analyse the capacity of a single quantum link between two devices where one device communicates periodically with data-bursts, separated by idle periods of fixed duration. Here we require no network routing and focus just on the single link. This type of data traffic profile is commonly found in Internet of Things (IoT) networks, specifically sensor networks\cite{dias2016iot}. To simulate this, we program the sender to perform periodic bursts of data packet transmissions to the receiver using superdense coding and then transmit EPR frames between data bursts, as described in the protocols in the previous section. The output statistics of the simulation provide a count of the occurrences of the various simulation events, such as the number of transmissions made over the fibre and the number of EPR pairs generated. We accumulate these statistics for various simulation parameters and show the results in  Fig.~\ref{fig:data_1}.

First, we simulated the effects of varying the number of qubits an EPR frame is composed of, with a fixed number of 10~EPR frames sent between bursts. The number of data packets in a single burst is also a simulation variable, where each data packet is composed of 168 classical bits. The upper plot in Fig.~\ref{fig:data_1} shows the trends of the average amount of data sent per transmission for varying both the length of the EPR frame as well as the number of packets in a burst. The trends show that---as one could expect---when fewer data packets are sent in a single burst, and an EPR frame is composed of more EPR pairs, the number of bits per transmission will more rapidly approach two. In the next simulation scenario, we simulated the transmission of data packets where a varying number of EPR frames generated in the idle times also considering EPR frames of fixed length. The results are depicted in the lower plot of Fig.~\ref{fig:data_1}. As expected, when there is more time to generate EPR frames between data transmissions, the average capacity of the channel increases.  

To reconfirm these results, by exploiting the periodicity in the setup, we can find analytic expressions matching our simulation as follows: Let $B$ be the number of packets in a burst, $E$ the EPR frames per burst, $D$ the bits in the packet, and $L$ the EPR pairs in an EPR frame. Then, the transmission time of one data burst is calculated to equal $EL + (DB - 2EL)$ whenever $EL<BD/2$ and the number of bits per transmission is $C(B,E,D,L)=BD/(DB - EL)$ when $EL < BD/2$, and $C(B,E,D,L)=2$ otherwise. We add the analytical result to the plots with a dashed lined and we see all that data points very closely matching the trends.

\section{Conclusions and Outlook}
    
In summary, we have demonstrated an integration of the classical network emulator ComNetsEmu with a quantum link layer, simulated using QuNetSim. As a proof-of-concept, we presented a classical-quantum link-layer protocol for transmitting classical information over a quantum network at an accelerated data rate, using pre-established quantum entanglement, where entanglement is generated during idle times in the network under a bursty network traffic model. 

Future work will consider other scenarios where data is transmitted in a more continuous stream. Using this network emulator, we plan to take the proposed encoding schemes of \cite{wilde2012}, which uses a \enquote{trade-off coding} approach, combining classical data with entanglement assisted communication in a single transmission, achieving higher communication rates in some cases. In \cite{pereg2021quantum}, there is another approach where only the receivers of the classical transmission share entanglement, which is used for message decoding. Using our simulation tool, we plan to test explicit protocols for each setting.

Because of the software architecture of QuNetSim, future work will also allow us to increase the level of realism in the quantum simulation, and, moreover to run such protocols on physical hardware without much change to the simulation code, as these synchronization tasks are abstracted away in QuNetSim. Overall, integrating quantum features into classical network emulators paves the way for classical network engineers to investigate their own use cases in a familiar setting but with the additional features offered by quantum networks, creating a tool for testing novel applications on quantum-enhanced networks.

\section*{Acknowledgements}
    
This work is partially funded by the Deutsche Forschungsgemeinschaft (DFG) as part of Germany’s Excellence Strategy–EXC2050/1–Project ID 390696704–Cluster of Excellence “Centre for Tactile Internet with Human-in-the-Loop” (CeTI) of TU Dresden, and via the Emmy-Noether grant NO-1129/2 (SD, JN, SS). HB, CD and RF were supported by the Bundesministerium f\"ur Bildung und Forschung (BMBF) through Grants 16KIS0856 (CD, RF), 16KIS0858 (HB), and 16KIS0948 (HB).
    
\bibliographystyle{IEEEtran}
\small{}

\end{document}